# OHMF: A Query Based Optimal Healthcare Medication Framework

Santosh Kumar Majhi[1], Padmalochan Bera[1]
[1]School of Electrical Sciences
Indian Institute of Technology, Bhubaneswar, 751013 INDIA
Email: {sm20, plb}@iitbbs.ac.in

**Abstract-** Today, cloud computing infrastructure is largely being deployed in healthcare to access various healthcare services easily over the Internet on an as-needed basis. The main advantage of healthcare cloud is that it can be used as a tool for patients, medical professionals and insurance providers, to query and co-ordinate among medical departments, organizations and other healthcare related hubs. Although healthcare cloud services can enable better medication process with high responsiveness, but, the privacy and other requirements of the patients need to be ensured in the process. Patients' medical data may be required by the medical professionals, hospitals, diagnostic centers for analysis and diagnosis. However, data privacy and service quality cannot be compromised. In other words, there may exist various service providers corresponding to a specific healthcare service. The main challenge is to find the appropriate providers that comply best with patients' requirement.

In this paper, we propose a query based optimal medication framework to support the patients' healthcare service accessibility comprehensively with considerable response time. The framework accepts related healthcare queries in natural language through a comprehensive user-interface and then processes the input query through a first-order logic based evaluation engine and finds all possible services satisfying the requirements. First order logic is used for modeling of user requirements and queries. The query evaluation engine is built using zChaff, a Boolean logic satisfiability solver. The efficacy and usability of the framework is evaluated with initial case studies on synthetic and real life healthcare cloud.

## I. INTRODUCTION

Cloud Computing is a model for enabling ubiquitous, convenient, on-demand network access to a shared pool of configurable computing resources (e.g. networks, servers, storage, applications and services) that can be rapidly provisioned and released with minimal management effort or service provider interaction. Recent Research by Gartner [1], on top 10 "disruptive technologies," outlined that enterprise cloud [2], power grids [3], web ecosystems, virtualization and social software are the dominant but threat-prone technologies which are being adopted largely in different countries.

Today, healthcare is undergoing enormous change and reform worldwide. With healthcare spending as a percentage of GDP rising by double digits annually in some countries [4] [5] [6], and concerns growing over medical care access and quality, governments and healthcare institutions are working to find creative new ways to address the need for improved care delivery models and payment reform.

Cloud computing [7] [8] provides efficient and cost-effective way of delivering IT services which can directly support the need to slow the growth of healthcare costs. The agility provided by on-demand, flexible cloud-based computing resources can also help empower the growth of a new generation of healthcare services and initiatives that respond more quickly and creatively to the needs of people and organizations across the continuum of health. Many healthcare providers and insurance companies today have adopted some form of electronic medical record systems, though most of them store electronic health record (EHR) [4] of patients in centralized databases. Typically, a patient may have many healthcare providers, including primary care physicians, specialists, therapists, and other medical practitioners. In addition, a patient may use multiple healthcare insurance companies for different types of insurances, such as medical, dental, vision, and so forth.

While cloud computing promises significant benefits, legitimate security and compliance concerns [9] have slowed cloud implementation within the healthcare domain, particularly due to multiple layers of statutory and regulatory requirements that govern the handling of protected health information. With the widespread use of electronic health record (EHR) [4], building a secure EHR sharing environment has attracted a lot of attention in both healthcare industry and academic community.

One unique concept in healthcare clouds is "patient-centric" view [10], which is a term used mostly in community healthcare systems. Community healthcare system [10] offers an open platform for patient to collect, store, use, and share health information in a controlled manner with ubiquitous accessibility. It also offers secure storage and management of patients' EHRs for multiple applications (e. g. disease treatment, lab research, insurance, and other social-networking applications). Most of the community healthcare cloud service models, such as Microsoft HealthVault [11] and Google Health [12], adopt a centralized architecture with patient-centric views. By patient-centric, it means that the information stored in the community EHR system is imported by patients and only can be made available to a variety of applications under the control of patients.

Research on the various security issues surrounding healthcare information systems has been heated over the last few years. The common security issues shared by healthcare cloud applications are ownership of information, authenticity, authentication, non-repudiation, patient consent and authorization, integrity and confidentiality of data. ISO/TS 18308 standard gives the definitions of security and privacy issue for EHR [4]. The Working Group 4 of International Medical Informatics Association (IMIA) was set up to investigate the issues of data protection and security within the healthcare environment. Its work to date has mainly concentrated on security in EHR networked systems and common security solutions for communicating patient data [13]. The European AIM/SEISMED (Advanced Informatics in Medicine/Secure Environment for Information Systems in Medicine) project is initiated to address a wide spectrum of security issues within healthcare and provides practical guidelines for secure healthcare establishment [14].

Across the life cycle of healthcare cloud service, a service can be provided through a chain of service providers. When a user starts searching for a specific service, he/she may find a large number of potential service providers. Therefore, the most challenging problem in cloud service provisioning is to find service providers that ensure the user's data is handled as agreed by the participating parties complying with their requirements. This is especially critical among the dynamic environment of the cloud. In the context of the Optimized health care framework has been given much emphasis leading to the development of significant number cloud interactive systems.

In this paper, we present a query based optimal medication framework to support the patients' healthcare service accessibility comprehensively with considerable response time.

This paper is organized is as follows. Section I describes the basics of cloud computing and its application in healthcare services. The related study on different healthcare cloud systems have been presented in section II. Section III describes the problem with a motivating example. In section IV, we present our proposed query based optimal healthcare framework for supporting the patients' healthcare service accessibility. Section V presents the evaluation of framework with some case studies. Finally, we conclude with scope of future work in section VI.

## II. RELATED WORK

A significant number of works have been initiated on healthcare cloud service management. Sabah Mohammed and et.al has discussed a distributed Web interactive system for sharing health records on the cloud using distributed OSGi services and consumers. This system(HCX (Health Cloud eXchange)) allows for different health record and related healthcare services to be dynamically discovered and interactively used by client programs running within a private cloud. A basic prototype was represented as the proof of concept along with a description to the steps and processes involved in setting up the underlying infrastructure. They showed how to build and integrate a composite application using the Eucalyptus and Apache CXF DOSGi open source frameworks for sharing CCR EHR records. The developed HCX prototype comprising of composite modules (distributed across the cloud) and can be integrated and function as a single unit. HCX allows adaptors and bridges to be created for existing EHR systems and repositories so that records can be exchanged through a standard interface and CCR record format. This is accomplished by building DOSGi based services and consumers made scalable through the cloud [15].

Ming Li et.al has presented the design and implementation of Personal Health Records and provided security to them while they are stored at third party such as cloud. This web based application that allows people to access and co-ordinate their lifelong health information. The patients have control over access to their own PHR. The framework addresses the unique challenges brought by multiple PHR owners and users, they have reduced the complexity of key management [16].

Leslie S. Liu et. al used a multi-method approach to evaluate PHR systems[17]. They have conducted various interview with potential end users like clinicians and patients and conducted evaluations with patients and caregivers as well as a heuristic evaluation with HCI experts. They focused on three PHR systems: Google Health, Microsoft HealthVault, and WorldMedCard. The results show that both usability concerns and socio-cultural influences are barriers to PHR adoption and use. They have presented those results as well as reflect on how both PHR designers and developers might address these issues now and throughout the design cycle.

Vassiliki Koufi et. al discussed the development of a PHR-based EMS in a cloud computing environment. Cloud-based services can prove important in emergency care delivery since they can enable easy and immediate access to patient data from anywhere and via almost any device[18].

DICOM-based system is based on Digital Imaging and Communications in Medicine to deal with the high volume of medical images and diagnostic imaging procedure [19]. DICOM server, Web User Interface, and DICOM file indexer are major components of DICOM. User can upload, search and browse images via the Web User Interface. The DICOM file indexer parses the header of DICOM files as they are uploaded from the client through the DICOM server or Web User Interface. The DICOM system design includes a portable application program interface layer which enables this open source service to be ported to different cloud platforms. It was tested with various public domain images. This prototype's system design and implementation shows the feasibility of using cloud computing to provide a long term offsite medical image archive solution. In the future work, the authors plan to implement more security using Internet protocol filtering and also evaluate complete compliance of Health Insurance Portability and Accountability Act (HIPAA).

Health ATM kiosks are developed for patients to manage their own personal health data [20]. It integrates services from Google's cloud computing environment. It provides timely access to relevant health data to patients and strengthens patients' communication with their care providers. Individuals can review personal account information and perform transactions to manage their care online. This functionality is particularly helpful for patients with chronic conditions who need to monitor their health on a daily basis. They mainly serve non-native English speakers, undocumented individuals, displaced individuals, elderly persons, migrant farm workers and homeless individuals. It is also a cost effective solution of personal healthcare management. A Health ATM kiosk makes use of cloud computing architectures, both vertically and horizontally. Horizontally, it can connect and integrate multiple clouds to function as one logical cloud; while vertically, it can improve capacity of a cloud by enhancing individual existing nodes in the cloud. Currently, the systems cannot be directly handed over to patients. In order for Health ATM kiosks to be more effective, constant training, outreach, education and collaboration are must. A further plan is to incorporate feeds from different health-related services in order to include data available from pharmacies.

The scenario in which patients and health care providers log into the front-end of a web services to give input all of patient's information, including medications, allergies, laboratory results, medical images and so on, from wherever that patient has received medical care might be envisioned for healthcare environments of the future by evolving toward the cloud [21][22].

From the above literature study, we found that there is no considerable attempt made in integrating various healthcare service providers and in providing comprehensive systems for the patients' to access and evaluate those service providers based on their requirements. In this paper, we propose a query based optimal medication framework to support the patients' healthcare service accessibility comprehensively with considerable response time.

### III. MOTIVATING EXAMPLE

In this section, we demonstrate the problem with a motivating example.

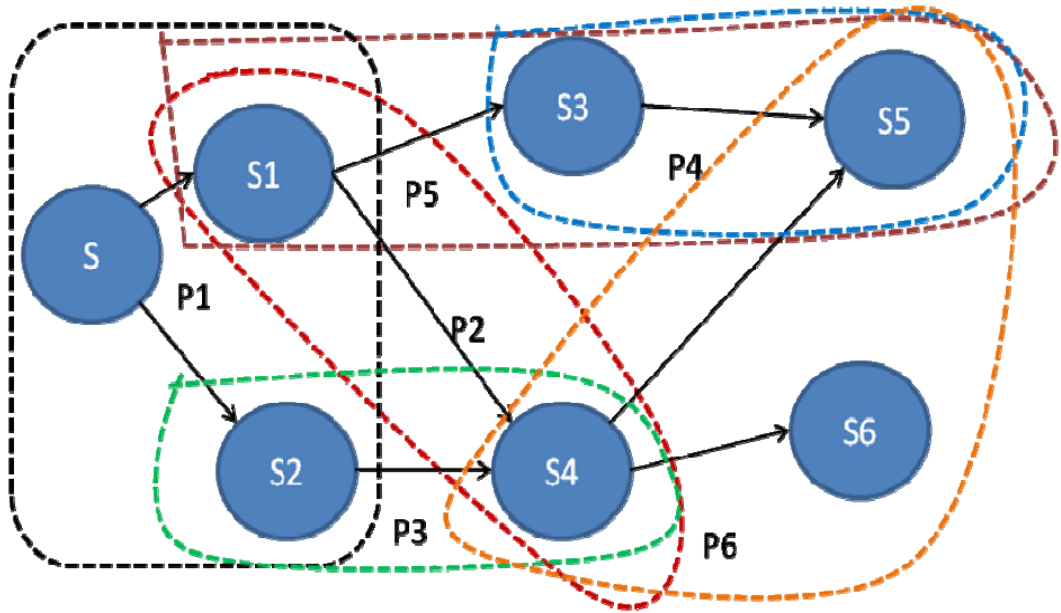

Figure 1. A simple Service Chain for the given example

Across the life cycle of healthcare cloud service, a service can be provided through a chain of service providers. In such scenario, the most important question is: "how can an user find the service providers that comply with his/her requirements in terms of service quality and privacy?"

Let us consider that a user requests service S for admitting a patient in hospital for preventive medication. Now, the selection of hospitals depends on satisfaction of patient's privacy and other service requirements. A hospital may deploy doctors for specific medication process. It may also require specific tests to be done through other diagnostic centers. The related insurance claims are processed by various agencies. It is very obvious that the service S is served as a chain of service providers [S1, S2] (hospitals), [S3, S4] (diagnostic centers), and S5 (enrolled insurance organization) and S6 (floating insurance organization). The patient can enroll with the floating insurance organizations only if the corresponding hospital is tied-up with that insurance organization, may involve many more. Now, the privacy, data access and service related constraints can be stated as follows:

(i) The hospitals must maintain the privacy of the patient to protect the specified records. While, transferring data electronically, hospital should have appropriate mechanism to encrypt the records and data should be transferred trough secured tunnel. Electronic transfer of the prescribed medicine information and test records is not allowed to insurance organizations. A hard copy of such documents with authorized signature may be allowed. (ii) The hospitals should be well equipped with related medical facilities and quality doctors (may be a specific list given by the patient). The hospital should have intensive care unit facility with high-end medical instruments.

(iii) Diagnostic Centers can only view specific details of a patient required for the related tests but not about the actual disease and patients other details. This requirement can be fine-grained based on the patients need. Such as, the specialist doctor in the diagnostic center may look at some history of patient for emergency assessment however the attendants or report generating people should not able to access those information. In addition, the people in the diagnostic center who are socially connected with the referring doctor should be prevented from accessing the records.

(iii) Insurance organization may need to look at the records of the medicines prescribed by the doctors and tests done for assessing the bill (as per the insurance policy). However, the reports and final discharge details shouldn't be disclosed to them.

In this scenario, the major challenge is to find the appropriate service providers that satisfy the patient's requirements. The complexity of this problem increases with dependency between the heterogeneous service providers with their policies and implementations. In summary, various cloud based medical applications and systems are studied. It has been observed from the related study and motivating example, that various healthcare systems are being developed with two major objectives (i) to improve the overall medication process from the perspective of patients, doctors and hospital networks (ii) to assist patients in managing their own care effectively especially in emergency situations. To the best of our knowledge, there exists a few cloud supported systems that provide information sharing between hospitals to guide the medication process. Also there is no such tool exist for extracting information regarding various features of hospitals and insurance providers that satisfy patients' requirements. This motivates us to propose a query based optimal healthcare framework for supporting the patients' healthcare service accessibility comprehensively with considerable response time.

## IV. PROPOSED FRAMEWORK

In this section, we describe our proposed query based optimal healthcare service framework. It allows automated searching of various medical information in a large hybrid health care cloud from different service providers. This will in turn guide the patients to take necessary and optimal decisions related to the specific healthcare service at any point of time. The framework is built on top of a formal query evaluation engine with trusted backend healthcare cloud. It has the following three major components:

(i) User Interface
(ii) Query translation and Evaluation Engine
(iii) Back end healthcare cloud (database + networks)

**(i) User Interface**

The proposed framework provides a comprehensive user interface through which user can place related queries with necessary requirements and get the response after evaluation of the query. For the purpose of evaluation of queries, the interface provides convenient connectivity to the Query Evaluation Engine and backend healthcare cloud. The GUI has two major components (i) User Validation Module (ii) Query Specification and Response Module.

The user validation module checks the user credentials using simple authentication scheme. It essentially checks the login password and the role of the user to validate his/her credential. On the other hand, the query specification and response module allows the authenticated users to place their query in natural language with specific syntax and semantics. The syntax of query essentially consists various attributes of healthcare services such as patient centered, clinical excellence, connection with trusted insurance policies, cost of medication, etc. The user can comprehensively specify their query as per the requirements. It is to be noted that our query specification module provides various tabs/interface icons to select corresponding attribute values. Figure 2 shows a generic view of the interface with various attributes. A query follows very simple semantics which expresses it as an expression. An expression can be conjunction (logical AND) of mandatory attribute fields connected with disjunction (logical OR) of optional attribute fields and negation of except attribute (logical NOT) fields. Each attribute can be also expressed as an expression. Mandatory, optional and except attribute set may vary from user to user depending on their requirements. The simple grammar corresponding to a query is represented as follows:

$$G(L): (\{E\}, \{\wedge, \vee, \neg, (,), A_i, \in\}, P, E)$$

Where, P indicates the set of production rules as follows:

$$E \rightarrow (E \wedge E)/(E \vee E)/E/\neg E$$
$$E \rightarrow A_1/A_2/.../A_i$$

An example of a query is presented as follows:

**Example:** Find the list of hospitals which provides 100% patient centered ($A_1$) service with specified clinical excellence ($A_2$) and 60% coverage by insurance policy ($A_3$); whereas there is no constraint on the expenses ($A_4$) of the overall medication process.

This query can be mapped to the following reduced grammar corresponding to the earlier representation.

$$E \rightarrow (E \wedge E)/\neg E$$
$$E \rightarrow A_1/A_2/A_3/A_4$$

The actual expression for the above query can be represented as follows:
$$E \rightarrow (A_1 \wedge A_2 \wedge A_3) \wedge \neg A_4$$

For the purpose of inputting queries and observing the responses, the GUI also includes of a query specification and response module.

**Query Specification and Response Module:** This module accepts the user specified queries and provides response of the queries according to the specified syntax and semantics. Our interface provides a split screen where the first split is used for specifying the query and second split is used for showing the response of the query. The query specification module detects the syntactic and semantic errors (if any) in the query and reports to the users. The user requirements will be captured and stored in a file. The stored query(in a file) will be communicated to the Query translation and Evaluation engine module.

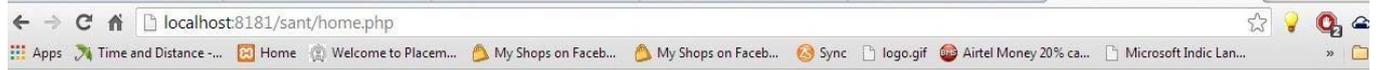

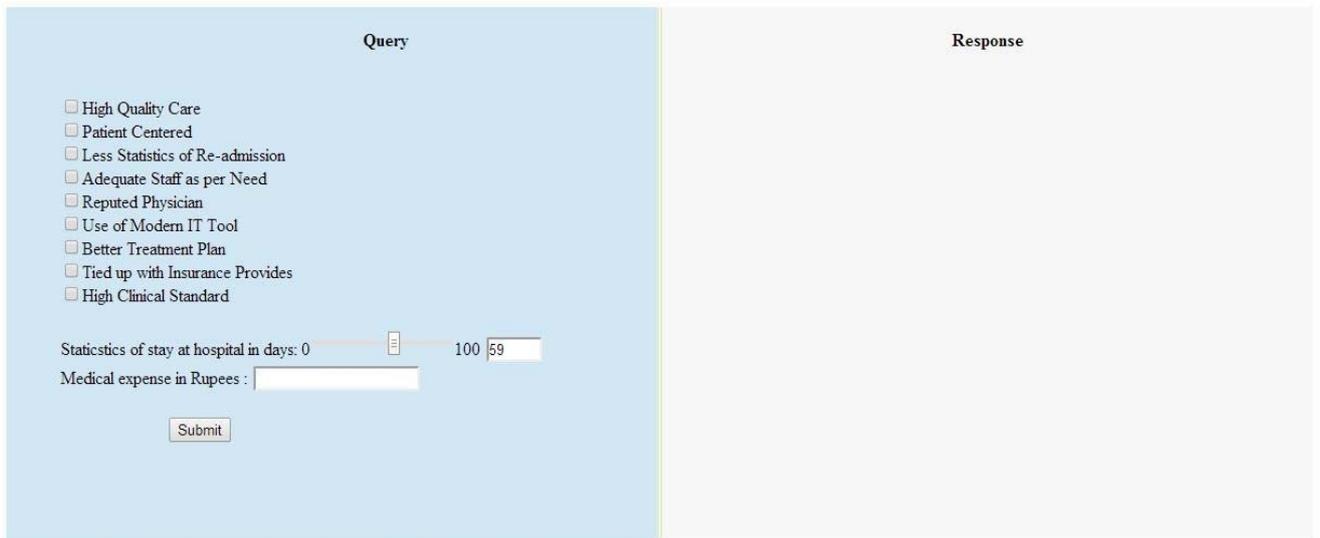

Figure 2. Comprehensive User Interface

### (ii) Query translation and Evaluation Engine

The stored query given by the query specification module will be translated to first order logic based expression and evaluated by the SAT solver supported engine. Then, this module will interact with the backend healthcare database to check the satisfiability of the query with respect to the information stored in the database. For this evaluation, an image of the database will be derived in first order logic Boolean function. Finally, the results will be presented to the users based on the evaluation. The components of this module are (i) Query Translation (ii)SAT Solver (iii) Response Engine.

**Query Translation:**
This module converts the stored queries into first order logic query. The first order logic query is essentially represented in conjunctive normal form (CNF) to comply with the SAT solver. An example of a first order logic query is shown as follows:
List the hospitals with high quality care ($R_1$) or patient centered ($R_2$), low cost($R_3$) and not tied up with good insurance agency($R_4$).

The argument will be written as $(R_1 \vee R_2) \wedge R_3 \wedge (\neg R_4)$

The 3-SAT algorithms are used to convert the translated stored query to CNF form. Now the query will be mapped to CNF by using 3-SAT algorithm. The form of the query will be as follows:

$$P_{ri} = \bigvee_{i=1}^{m}(X_m) \quad \text{and} \quad Q = \bigwedge_{i=1}^{n} P_{ri}$$

This CNF form is used to do computation with SAT Solver.

We can write the user query is the function of user requirements. So, $Q :\rightarrow f(a_1, a_2, a_3, ...)$

**SAT Solver:** This module represents the image of the backend cloud into first order logic function. Here the database is a function of various user requirements(i.e., attributes) stored with the backend server. We refer to the corresponding first order logic database cloud model as $M_c$.

$M_c = f(a_1, a_2, a_3, ...)$

Now the evaluation engine checks the satisfiability of the translated query Q with the cloud models. So, $Q \rightarrow M$ represents the cloud model corresponding to user query. Sometimes the particular user query does not satisfy or mapped to the exact available cloud models. In that case, if we change some requirements partially, it will give some result. The SAT solver will give the possible traces of solution as per the requirement. Then the result trace will be foreword to Response Engine.

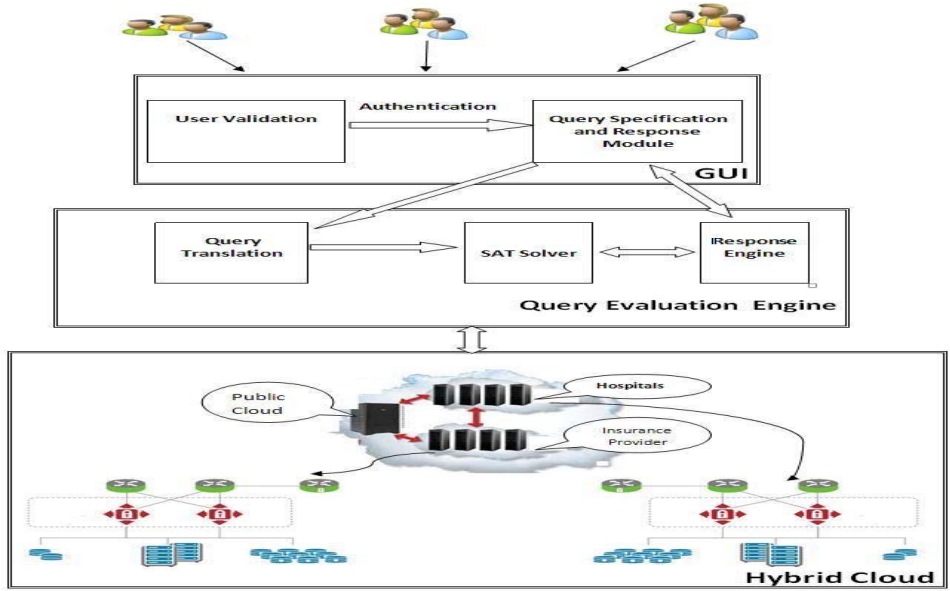

Figure 3. Architecture of the proposed optimal medication system

**Response Engine:** The output result will be translated into natural languages and the response engine will communicate it to the GUI.

**(iii)Back end healthcare cloud**

This module consists of different kind of clouds, which can be hospital clouds, insurance providers cloud, health care testing clouds and other computing resources. The role of this layer to interconnect various networks of private hospital clouds, different insurance provider public or private clouds and medical diagnosis testing clouds. The above clouds are may interconnected via public cloud and/or internet. In each of the above clouds includes boarder routers, core switching configuration, VLANs, Firewall, Load balancing, Reporting and configuration databases, Cloud servers, Configurable CPU,RAM and Storage, Cloud files and cloud based storage. It aims to provide enormous computing power.

## V. EVALUATION WITH CASE STUDY

The initial version of the framework is developed using PHP and MySQL server(XAMP) database. The query evaluation engine is built on top of zChaff Boolean SAT solver [23]. The translation of queries into corresponding SAT clauses is implemented using 3-SAT algorithm [24]. The framework is evaluated with few synthetic set of patient requirements under a specified healthcare cloud infrastructure. The case study considers the patient requirements as referred in Table 1. We have analyzed the results of the query evaluation engine in form of SAT instances. It has been observed that the query results are correct with respect to the healthcare cloud database.

| Sl.No. | Requirements attributes | Notation |
|--------|------------------------|----------|
| 1 | High quality care | $R_1$ |
| 2 | Patient centered | $R_2$ |
| 3 | Minimum length of stay at hospital | $R_3$ |
| 4 | Less statistics of readmission | $R_4$ |
| 5 | Adequate staff as per the need | $R_5$ |
| 6 | Low cost | $R_6$ |
| 7 | Available Medical Services | $R_7$ |
| 8 | Reputed Physician | $R_8$ |
| 9 | Clinical Standards | $R_9$ |
| 10 | Use of Modern IT tool | $R_{10}$ |
| 11 | Tied up with good Insurance Agency | $R_{11}$ |
| 12 | Better Treatment Plan | $R_{12}$ |

Table 1. List of Requirement attributes

The efficacy of the query evaluation engine is studied with respect to SAT translation/modeling time and SAT solver run time with varying query and database size. Query and database size indicates the cumulative sum of the associated range of values to each different attribute. For example, if there are 5 different attributes, e.g., $a_1$, $a_2$, $a_3$, $a_4$ and $a_5$ with their value ranges (1-100), the query size will be 500. However, in reality these ranges are heterogeneous. The SAT translation time includes translation of database and the query into Boolean models. For example, the attribute $a_1$ can be reduced into 7 Boolean variables $b_0$, $b_1$, $b_2$, $b_3$, $b_4$, $b_5$, $b_6$, and $b_7$. Thus, the requirement with attribute $a_1$ is represented as a Boolean clause, $f(b_0, b_1, .. b_7)$. A complete query is reduced into Boolean function $f_a(a_0, a_1, ..., a_n)$. In our evaluation, the query size is varied from 100 to 5000. The results are shown in Table 2 which highlights the query size, SAT translation time and SAT run time. It has been observed that the SAT translation time is linearly dependent on the size of the queries whereas the SAT run time almost remains constant. This is because of the fact that the SAT translation time involves deriving Boolean clauses corresponding to the query and the complexity of the query depends on the number of reduced Boolean variables. The number of reduced Boolean variables is linearly dependent on the query size. On the other hand, SAT run time almost remains constant because the zChaff SAT solver internally use guided search technique called Chaff [25] for finding solutions. It depends on the complexity and dependency between the Boolean clauses. This shows the power of the SAT solvers in solving Boolean formulae. Therefore, our framework is able to find the query result in considerably good time. It takes 14.91 msec to process a query with size 5000. At present, the translation of SAT results into natural language output is performed manually.

| Query Size | SAT translation time (msec) | SAT solver run time (msec) |
|---|---|---|
| 200 | 8.23 | 1.2 |
| 350 | 8.89 | 1.2 |
| 1000 | 9.32 | 1.25 |
| 1500 | 10.12 | 1.26 |
| 2300 | 11.25 | 1.29 |
| 3000 | 11.98 | 1.31 |
| 3500 | 12.37 | 1.32 |
| 4000 | 12.89 | 1.33 |
| 5000 | 13.56 | 1.35 |

Table 2. Results on query evaluation time

VI. CONCLUSION AND FUTURE WORK

This paper presents a comprehensive query based optimal healthcare service framework that can be targeted to automate the searching of various medical information in a large hybrid healthcare cloud from different service providers. This will in turn guide the patients to take necessary and optimal decisions related to the specific healthcare service at any point of time. The framework may facilitate the service providers to design a scalable and efficient patient interaction system to choose the services as per their requirements. The major process elements of the proposed framework are as follows:

- The user validation and requirement gathering has been done through a simple and comprehensive user interactive interface.
- Query Translation module translates the user specified requirements into first order logic query. It is represented in Boolean conjunctive normal form (CNF).
- The SAT solver checks the satisfiability of the translated query with the healthcare cloud backend database. For this, a Boolean model is extracted from the database.
- The SAT solver provides the possible traces of solution that satisfies the requirement. Then the result trace is provided as output through the Response Engine.
- The output is translated into natural language forms and the response engine communicates the same to the GUI.

The query evaluation engine is developed using zChaff SAT solver. The efficacy of the framework has been demonstrated through a case study. The proposed methodology has been shown to be correct and scalable. In future, the framework will be validated with various real life case studies. This work can be extended to develop an integrated patient interaction system tool for systematic retrieval of healthcare information.